\documentclass[lettersize,journal]{IEEEtran}
\usepackage{amsmath,amsfonts}
\usepackage{algorithmic}
\usepackage{algorithm}
\usepackage{array}
\usepackage[caption=false,font=normalsize,labelfont=sf,textfont=sf]{subfig}
\usepackage{textcomp}
\usepackage{stfloats}
\usepackage{url}
\usepackage{verbatim}
\usepackage{graphicx}
\usepackage{cite}
\usepackage{bm}
\usepackage{multirow}
\usepackage[normalem]{ulem}
\useunder{\uline}{\ul}{}
\usepackage{booktabs}
\usepackage{tipa}
\usepackage{diagbox}
\usepackage{footmisc}
\usepackage{hyperref}
\usepackage[hyphenbreaks]{breakurl}
\usepackage{xcolor}

\begin{document}

\title{Incorporating Ultrasound Tongue Images for Audio-Visual Speech Enhancement}

\author{Rui-Chen Zheng,~\IEEEmembership{Student Member,~IEEE},~Yang Ai,~\IEEEmembership{Member,~IEEE},~Zhen-Hua Ling,~\IEEEmembership{Senior Member,~IEEE}
\thanks{This work is the extended version of our conference paper \cite{zheng2023incorporating} accepted by INTERSPEECH 2023. }
\thanks{The authors are with the National Engineering Research Center of Speech and Language Information Processing, University of Science and Technology of China, Hefei 230027, China. Contact: zhengruichen@mail.ustc.edu.cn; yangai@ustc.edu.cn; zhling@ustc.edu.cn. This work was partially supported by the Fundamental Research Funds for the Central Universities under Grant WK2100000033.}
}



\maketitle

\begin{abstract}
Audio-visual speech enhancement (AV-SE) aims to enhance degraded speech along with extra visual information such as lip videos, and has been shown to be more effective than audio-only speech enhancement. 
This paper proposes the incorporation of ultrasound tongue images to improve the performance of lip-based AV-SE systems further.
To address the challenge of acquiring ultrasound tongue images during inference, we first propose to employ knowledge distillation during training to investigate the feasibility of leveraging tongue-related information without directly inputting ultrasound tongue images. Specifically, we guide an audio-lip speech enhancement student model to learn from a pre-trained audio-lip-tongue speech enhancement teacher model, thus transferring tongue-related knowledge.
To better model the alignment between the lip and tongue modalities, we further propose the introduction of a lip-tongue key-value memory network into the AV-SE model. This network enables the retrieval of tongue features based on readily available lip features, thereby assisting the subsequent speech enhancement task.
Experimental results demonstrate that both methods significantly improve the quality and intelligibility of the enhanced speech compared to traditional lip-based AV-SE baselines. 
Moreover, both proposed methods exhibit strong generalization performance on unseen speakers and in the presence of unseen noises. 
Furthermore, phone error rate (PER) analysis of automatic speech recognition (ASR) reveals that while all phonemes benefit from introducing ultrasound tongue images, palatal and velar consonants benefit most.

\end{abstract}


\section{Introduction}
\label{Section: Introduction}

\IEEEPARstart{S}{peech} enhancement (SE) is a critical research problem in speech signal processing, aiming to improve the quality and intelligibility of speech signals corrupted by various types of distortions \cite{kumar2016speech}. SE has wide-ranging applications in various fields, such as ensuring clear telecommunications in noisy environments, enhancing speech quality for hearing aids, improving the robustness of speech recognition systems, and so on\cite{chaudhari2015review}. With the increasing demand for high-quality speech in various applications, SE has become an active area of research, with many advanced algorithms and techniques being proposed.

Classical audio-only speech enhancement (AO-SE) approaches have been successful in estimating underlying target speech signals. Early AO-SE methods \cite{ephraim1995signal, yang2005spectral, wang2006computational, loizou2013speech} rely on assumptions about the statistical characteristics of the involved signals and aim to estimate the target speech signals using mathematically tractable criteria \cite{michelsanti2021overview}. More recent AO-SE methods \cite{xu2013experimental, xu2014regression, pascual2017segan, defossez2020real} based on deep learning have shifted away from this knowledge-based approach towards a data-driven paradigm. These methods typically treat SE as a supervised learning problem, leveraging large amounts of labeled data to train deep neural networks (DNNs) capable of directly mapping noisy speech signals to their corresponding clean counterparts.

Though AO-SE methods have been widely used for SE, their performance can be limited in some scenarios, especially at extremely low signal-to-noise ratios (SNRs) and when the noise is highly correlated with the speech signal. Since speech perception is inherently multi-modal, particularly audio-visual, recent
studies have investigated using visual information in addition to acoustic signals for SE \cite{michelsanti2021overview}. This approach, known as audio-visual speech enhancement (AV-SE) \cite{almajai2010visually, yang2022audio, gabbay2018visual, hegde2021visual, hussain2021towards}, has shown to be more effective than simple AO-SE methods. The basic idea is that as visual information is essentially unaffected by the acoustic environment, it can provide complementary cues to the acoustic information to assist with SE. Besides, it can also help disambiguate phonetically similar sounds since it usually records the movement of articulators involved in speech production. 

Lip videos, which can be readily acquired using a camera, are the most commonly used visual cues for AV-SE due to their easy availability. Several recent studies have proposed deep learning-based AV-SE models using lip information. For example, a deep AV-SE model based on convolutional neural networks (CNNs) \cite{afouras2018conversation} was proposed to separate a speaker's voice by predicting both the magnitude and the phase of the target signal given lip regions. A novel framework \cite{xu2022vsegan} incorporated lip information for speech enhancement by integrating a generative adversarial network (GAN) to generate high-quality clean speech. Another study \cite{xu2022improving} proposed an AV-SE system that achieved impressive performance using a multi-layer fusion model with a multi-head cross-attention mechanism to fuse audio and lip features. These audio-lip SE models demonstrate stronger abilities to improve speech quality and intelligibility than AO-SE methods through the integration of lip information.

Speech production is a complex process relying on multiple articulators, including the jaw, lips, teeth, and tongue. Only using lip videos for speech processing tasks like SE often has limitations due to the lack of descriptions on internal articulators, e.g., tongue and velar. Some studies have suggested employing internal articulation features captured using medical imaging techniques such as magnetic resonance imaging (MRI) \cite{scott2014speech} and ultrasound tongue imaging (UTI) \cite{stone2005guide}, to provide complementary data. Compared to MRI, UTI is relatively cheap, non-invasive, and can provide high-resolution images. UTI uses a real-time B-mode ultrasound transducer placed under the speaker’s chin to visualize a midsaggital or coronal view of the tongue during speech production. Ultrasound tongue images have been used in various speech processing tasks, such as speech recognition \cite{liu2016comparison, xu2017convolutional, ribeiro2019speaker, tatulli2017feature, ji2018updating, ribeiro2021silent, wang2021improving} and speech reconstruction \cite{csapo2017dnn, toth2018multi, kimura2019sottovoce, porras2019dnn, zhang2021talnet, zheng2023speech}. However, their potential for SE tasks has yet to be fully explored, creating a research gap that could be addressed by leveraging the advantages of ultrasound tongue images. 

Therefore, this paper proposes incorporating ultrasound tongue images for traditional lip-based AV-SE systems to provide additional internal articulatory information, especially tongue information, to complement the external articulatory information provided by lip videos. To achieve this, we first propose developing an audio-lip-tongue SE model with a U-Net-based \cite{ronneberger2015u} structure, which utilizes both ultrasound tongue images and lip videos to assist SE. However, obtaining ultrasound tongue images during inference is more challenging than collecting lip videos due to the requirement of extra equipment. To enable AV-SE model to leverage the tongue knowledge even when ultrasound tongue images are unavailable during inference, this paper further proposes the following methods:
\begin{enumerate}
    \item The first proposed method, namely knowledge distillation-based (KD-based) audio-lip SE method, incorporates tongue information into a lip-based AV-SE model through knowledge distillation (KD). Specifically, an audio-lip SE student model is trained to learn tongue information from the pre-trained audio-lip-tongue SE model serving as the teacher via multiple loss functions. This facilitates the transfer of tongue knowledge from teacher to student model.
    \item The second method, namely memory-based audio-lip SE method, introduces a lip-tongue key-value memory network to the encoder of the audio-lip-tongue SE model. During training, all three modalities (noisy speech, lip videos, and ultrasound tongue images) are input, and the tongue features derived from ultrasound tongue images are stored in the memory spontaneously while the whole model are trained to perform SE. At inference time, the stored tongue features can be retrieved according to the input lip modality and then used to aid the subsequent SE task.
\end{enumerate}

The two proposed audio-lip SE methods only require the input of noisy audio and lip videos during inference, eliminating the need for inputting ultrasound tongue images while effectively utilizing tongue information.
The first method employs the KD framework directly on the three-modality audio-lip-tongue SE model to verify the feasibility of two-modality input during inference. It offers the advantage of achieving improved performance without requiring additional parameters for the traditional audio-lip SE mode. 
The second method is a further improvement over the KD-based method by explicitly modeling the alignment between lip and tongue modalities, specifically addressing its limitation of implicitly fusing lip and tongue features while transferring tongue knowledge. This enhancement significantly narrows the performance gap between the audio-lip SE model and audio-lip-tongue SE model, and even outperforms the latter in certain metrics.
Experimental results demonstrate that both proposed methods can generate speech with improved quality and intelligibility compared to conventional lip-based AV-SE baselines trained only with clean speech supervision. In addition, both proposed methods exhibit strong generalization performance on unseen speakers and in the presence of unseen noises. Moreover, a notable reduction in phoneme error rate (PER), especially for palatal and velar consonants, can be witnessed while applying automatic speech recognition (ASR) for transcription.

The remainder of this paper is organized as follows. Section \ref{Section: Related Work} reviews related work on traditional lip-based AV-SE systems and other speech processing tasks with ultrasound tongue images as input. Section \ref{Section: Proposed Methods} describes the three-modality audio-lip-tongue SE model along with the KD-based and memory-based audio-lip SE methods  in details. Section \ref{Section: Implementation Details} presents the implementation details of our proposed methods and Section \ref{Section: Experimental Results} presents our experimental results. Finally, Section \ref{Section: Conclusion} provides the concluding remarks.

\section{Related Work}
\label{Section: Related Work}

\subsection{Lip-Based AV-SE Methods}
Lip-based AV-SE methods have recently garnered substantial interest due to their potential in overcoming the limitations of AO-SE methods by combining the complementary articulation information provided by easily acquired lip videos with acoustic signals. This section presents a preliminary review of several lip-based AV-SE methods based on deep learning.

A common approach to perform AV-SE is by direct mapping, which leverages the audio and visual modalities to generate a denoised spectrogram directly. 
Hou et al. \cite{hou2018audio} demonstrated superior performance by integrating audio and visual streams through convolutional neural networks (CNNs), jointly generating enhanced speech and reconstructed images.
Adeel et al. \cite{adeel2019lip} proposed a lip-reading-driven deep learning framework for SE that combines deep learning and analytical acoustic modeling. Their approach utilized a stacked long short-term memory (LSTM)-based lip-reading regression model and an enhanced visually-derived Wiener filter to estimate clean audio features. 
Tan et al. \cite{tan2020audio} presented a multi-modal network for AV-SE, employing a two-stage strategy with separate modules for noise attenuation and dereverberation, trained using a novel multi-objective loss function.

Other AV-SE methods fall into the camp of spectrogram masking approaches, producing a spectrogram mask to mask the noisy audio spectrogram and transforming it back into the time domain to generate a clean waveform.
Gogate et al. \cite{gogate2018dnn} proposed a DNN-based audio-visual mask estimation model that integrates the temporal dynamics of audio and noise-immune visual features through stacked LSTM and convolution LSTM networks. 
Wang et al. \cite{wang2020robust} introduced a safe AV-SE approach using power binary masks (PBMs) to roughly represent speech signals, integrating visual-derived PBMs with audio-derived masks through a gating network.
Gao et al. \cite{gao20192, gao2021visualvoice} integrated visual information in the bottleneck layer of a U-Net model and output a mask for a complex spectrogram.

Researchers have also explored alternative techniques \cite{gabbay2018seeing, khan2018using, gu2020multi} for AV-SE tasks. Previous research \cite{michelsanti2019training} has demonstrated the superior performance of spectrogram masking regarding speech quality and intelligibility. Hence, in this paper, we construct a U-Net-based \cite{ronneberger2015u} model that generates a mask for a complex spectrogram as the foundational model, incorporating modality fusion at each encoder layer.

\subsection{UTI Incorporated Speech Processing Tasks}
Many researchers in the field of audio-visual speech processing have utilized UTI to capture internal articulatory features, recognizing the crucial role of internal articulators in both speech production and perception. This section reviews studies that have employed ultrasound tongue images as a visual modality for audio-visual speech recognition and speech synthesis. Within these studies, some are specifically centered around ultrasound tongue images, while others employ these images as additional articulatory inputs alongside lip videos.

\subsubsection{Speech Recognition} 
Previous studies \cite{hu23b_interspeech, liu2016comparison, xu2017convolutional, ribeiro2019speaker, tatulli2017feature, ji2018updating, ribeiro2021silent, wang2021improving} have explored the use of ultrasound tongue images for speech recognition. 
Some of these studies utilized ultrasound tongue images as an additional input to audio input in automatic speech recognition (ASR) systems. For example, Hu et al. \cite{hu23b_interspeech} introduced UTI-based articulatory movement features generated from parallel audio with an acoustic-to-articulatory inversion (AAI) model into the elderly and dysarthric speech recognition system, achieving better performance compared to the baseline ASR system that solely relies on acoustic features.
Other studies utilized ultrasound tongue images as an additional articulatory input to lip videos in silent speech recognition (SSR) systems.
Tatulli et al. \cite{tatulli2017feature} focused on continuous speech recognition using raw ultrasound tongue and lip videos, employing CNNs to extract visual features from each modality, which were then combined with an HMM-GMM decoder. 
Ribeiro et al. analyzed phonetic segment classification from ultrasound tongue images in different training scenarios \cite{ribeiro2019speaker}, and further investigated multi-speaker speech recognition using ultrasound tongue and lip video images for both silent and modal speech \cite{ribeiro2021silent}. 
Wang et al. \cite{wang2021improving} proposed a 3D CNN architecture that predicted future frames in ultrasound tongue and lip images, generating features for continuous HMM-based speech recognition.

\subsubsection{Speech Reconstruction} 
Ultrasound tongue images have also been proposed for speech reconstruction in previous studies \cite{csapo2017dnn, toth2018multi, kimura2019sottovoce, porras2019dnn, zhang2021talnet, zheng2023speech}, which aims to regenerate speech corresponding to the provided articulatory input. Despite sharing the common goal of generating clean speech with speech enhancement, it distinguishes itself by excluding noisy audio as part of its input. Some studies explore the performance of reconstructing speech solely from ultrasound tongue images. For example, 
Csap{\'o} et al. \cite{csapo2017dnn} employed ultrasound tongue images and speech signals as input with DNNs to estimate MGC-LSP coefficients for speech synthesis. 
Toth et al. \cite{toth2018multi} improved speech synthesis by using multi-task training for simultaneous speech recognition and synthesis as a weight initialization step. 
Kimura et al. \cite{kimura2019sottovoce} extracted acoustic features from a sequence of ultrasound images, leveraging internal information observed by an ultrasound imaging sensor to reconstruct speech. 
There are also studies using ultrasound tongue images as additional articulatory input to lip videos.
Zhang et al. \cite{zhang2021talnet} introduced an encoder-decoder model for speech reconstruction with ultrasound tongue and lip videos as inputs, consisting of separate encoders for processing tongue and lip data streams. 

The use of ultrasound tongue images in speech recognition and reconstruction highlights their valuable intra-oral articulator information for speech processing. However, their potential in SE tasks remains unexplored. This paper proposes integrating ultrasound tongue images to complement external articulatory information from lip videos in lip-based AV-SE systems, thereby providing internal articulatory details, particularly pertaining to the tongue. 

Furthermore, obtaining ultrasound tongue images during inference presents greater challenges compared to collecting lip videos due to the requirement for additional equipment. While Hu et al. \cite{hu23b_interspeech} utilized an AAI model to generate ultrasound tongue features corresponding to clean speech, its applicability to the AV-SE scenario is challenging as only noisy speech is provided during inference. Remarkably, the majority of current studies on speech recognition or reconstruction that integrate ultrasound tongue images alongside lip videos \cite{tatulli2017feature, ribeiro2021silent, zhang2021talnet} do not address the challenge of limited availability of ultrasound tongue images during practical inference. In response to this challenge, this paper proposes two methods designed to avoid the acquisition of tongue features with the assistance of lip videos during inference.

\section{Proposed Methods}
\label{Section: Proposed Methods}
This section presents our proposed methods for integrating ultrasound tongue images into AV-SE. We begin by introducing our proposed audio-lip-tongue SE model. It is based on spectrogram masking with a U-Net \cite{ronneberger2015u} style architecture, taking three modalities as input: noisy audio, lip videos, and ultrasound tongue images. 
However, acquiring ultrasound tongue images during inference is often challenging and costly. To address this issue, we propose two methods: KD-based and memory-based audio-lip SE methods. These approaches allow the use of valuable tongue information for AV-SE even in the absence of ultrasound tongue images during inference. 
While the KD-based method demonstrates the initial feasibility of employing only two modalities for inference, the memory-based method further overcomes the limitations of the KD-based method and offers improvements. 
We next provide detailed descriptions of our proposed methods.

\begin{figure*}[t]
  \centering
  \includegraphics[width=0.99\linewidth]{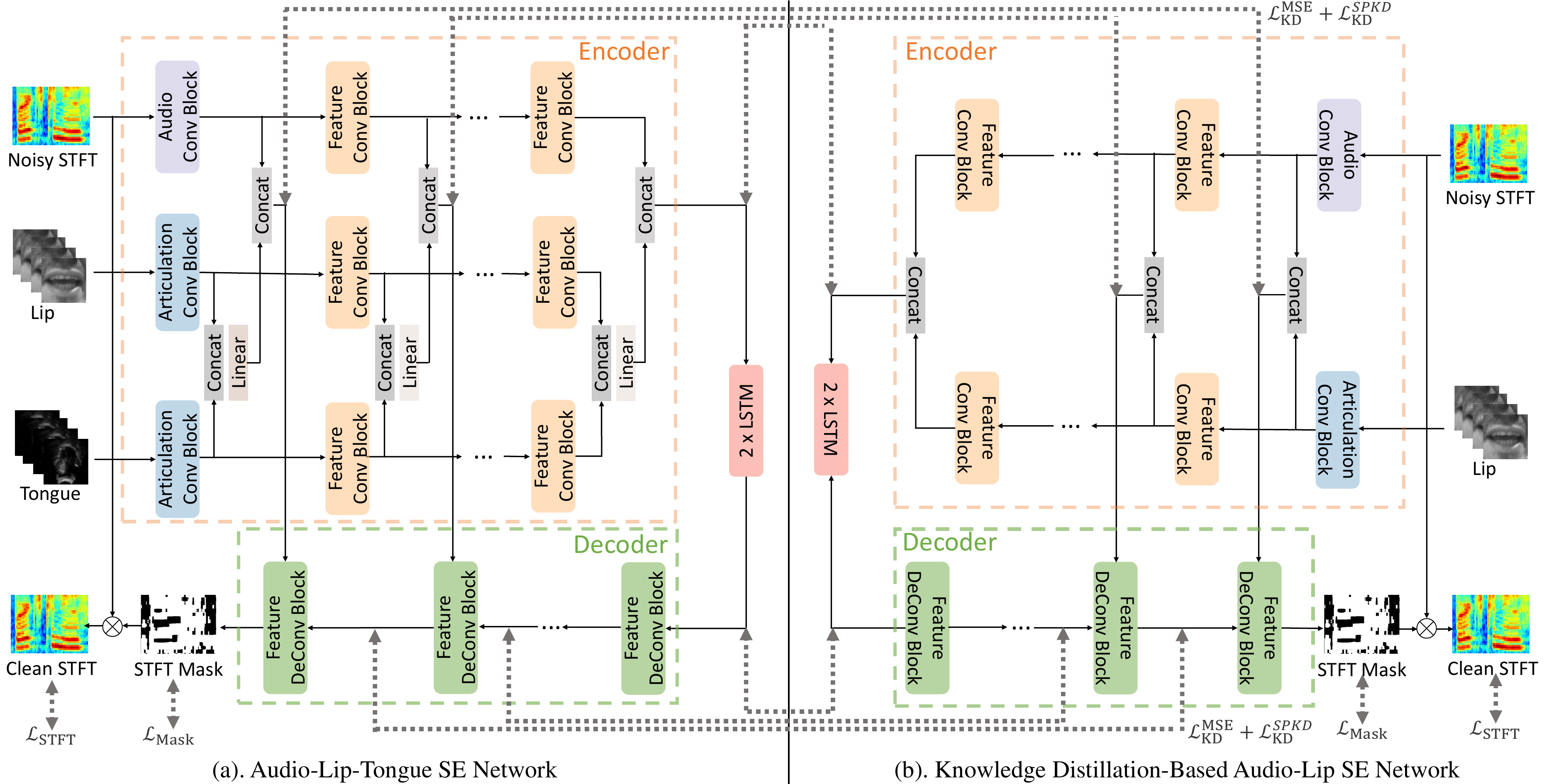}
  \caption{Details of the proposed audio-lip-tongue SE model and KD-based audio-lip SE method. The dashed arrows indicate the loss functions. (a) shows the structure of the proposed audio-lip-tongue SE model; (b) shows the structure of the audio-lip SE student model introducing tongue knowledge through the proposed KD method. }
  \label{Figure: Teacher and Student}
\end{figure*}

\subsection{Audio-Lip-Tongue SE Model}
\label{Section: Audio-Lip-Tongue SE Model}

Inspired by previous studies \cite{gao2021visualvoice, xu2022vsegan}, we propose a U-Net \cite{ronneberger2015u} style audio-lip-tongue SE model with noisy speech, lip videos, and ultrasound tongue images as input as shown in Fig.\ref{Figure: Teacher and Student}(a). The model can be roughly divided into a multi-modal encoder and a decoder connected by an LSTM-based embedding block.

The encoder contains an articulation stream and an audio stream. For the articulation stream, we process both lip and tongue image sequences to obtain fused representations for describing articulation. Specifically, pixel-wise mean and standard deviation are computed for each utterance, repeated, and then appended as extra channels to the ultrasound and lip sequences \cite{zhang2021talnet}. The resulting input is of dimension $3\times S\times H\times W$, where $S$ denotes the time dimension, and $H$ and $W$ are the height and width of the lip and tongue images. The input is first processed by an articulation convolutional block consisting of three strided 3D-CNN layers, each followed by batch normalization \cite{ioffe2015batch} and Leaky-ReLU.  The resulting features are subsequently transformed by concatenating the last two dimensions into a shape of $C\times S'\times D$, and then processed by seven feature convolutional blocks. Each block involves a series of 2D-CNN layers with pooling layers to reduce the $D$ dimension while preserving the time dimension. The lip and tongue features are concatenated at each layer and then passed through a linear layer to obtain the fused articulation representations with reduced size.

For the audio stream, the encoder takes as input the real and imaginary parts of the noisy complex spectrogram with dimension $2\times F\times S$, where $F$ is the frequency dimension of the spectrogram. Each time-frequency bin contains the real and imaginary parts of the corresponding complex spectrogram value \cite{gao2021visualvoice}. The input is first processed by an audio convolutional block comprising two strided 2D-CNN layers and then by seven feature convolutional blocks as are used in the articulation stream. 
Finally, the audio representation is concatenated with the fused articulation representations layer by layer to obtain the ultimate multi-modal representations for SE.

Two LSTM layers are inserted between the encoder and decoder to better model temporal dependencies. The decoder with skip connection exhibits a symmetric structure concerning the encoder, whereby the convolutional layer is substituted with an upconvolutional layer, and the pooling layer is replaced by a upsampling layer. The input of each decoder layer is the concatenation of the output of the previous layer and the multi-modal representations given by the corresponding encoder layer. The final output feature map is fed through an activation layer to predict a complex mask with the same dimensions as the input noisy spectrogram. The resulting mask is applied to the noisy input through complex multiplication, yielding an enhanced complex spectrogram which is transformed back into the time domain via an inverse short-time Fourier transform (iSTFT). 

The training criterion of the audio-lip-tongue SE model is to minimize the following loss
\begin{equation}
    \mathcal{L}_{SE} = \mathcal{L}_{Mask} + \alpha \mathcal{L}_{STFT},
    \label{Loss: SE}
\end{equation}
where $L_{Mask}$ and $L_{STFT}$ denote the mean square error (MSE) losses of the mask and the complex spectrogram, respectively. The hyperparameter $\alpha$ is utilized to ensure both $L_{Mask}$ and $L_{STFT}$ are scaled to the similar magnitude.

\subsection{KD-based Audio-Lip SE Method}
\label{Section: KD-based}
Under most circumstances, only a noisy speech and its corresponding lip video can be obtained as reference data since acquiring ultrasound tongue images is not as straightforward as capturing lip video. Hence, we propose enabling an audio-lip SE student model to assimilate ultrasound tongue information from a pre-trained audio-lip-tongue SE teacher model through KD. 

An audio-lip SE student model is proposed with the architecture depicted in Fig.\ref{Figure: Teacher and Student}(b), where the input of tongue images is removed compared with the teacher. We employ KD to train the student to utilize tongue information even in the absence of ultrasound tongue images during reference. Specifically, in addition to the supervised training with the backbone losses $L_{SE}$ described in Eq.(\ref{Loss: SE}), the student model also receives supervision signals from the teacher. During forward inference of the student and teacher networks, the output of each layer is saved to compute the KD loss for each layer of the encoder, decoder, and LSTM embedding blocks. For each layer, the MSE loss between the output features of the teacher and the student is calculated following
\begin{equation}
    \mathcal{L}_{KD}^{MSE} = \sum_{k=1}^{K} ||\bm{F}_{tea}^k- \bm{F}_{stu}^k||_2^2,
\label{Loss: KD_MSE}
\end{equation}
where $K$ represents the number of layers, and $\bm{F}_{tea}^k$ and $\bm{F}_{stu}^k$ denote the output features of the $k^{th}$ layer of the teacher and student model, respectively. We guarantee that the feature dimensions of the corresponding layers in the teacher and student models are identical.

The similarity-preserving knowledge distillation (SPKD) loss \cite{tung2019similarity, cheng2022cross} is utilized to supervise the student model further. The SPKD loss aims to achieve dimensional compression and simultaneously transmit similarity information by computing pairwise similarity matrices. Given a mini-batch input, the feature map of the $k^{th}$ layer is represented as $\bm{F}^k \in \mathbb{R}^{b\times c\times s\times f}$, where $b$ is the batch size, $c$ is the number of output channels, $s$ is the number of frames, and $f$ is the dimension of the feature space. To account for potential interference of information from different frames, the features are first segmented on frame-level and then flattened into two dimensions. The transformed feature of the $j^{th}$ frame is denoted as $\bm{F}^{(k,j)} \in \mathbb{R}^{b\times f'}$, where $f'=c\times f$. The similarity matrice of each frame for the teacher $\bm{G}_T^{(k,j)}$ and the student $\bm{G}_S^{(k,j)}$  is calculated independently following \cite{cheng2022cross}
\begin{equation}
\label{Equation: Similarity Matrice}
\begin{aligned}
&\Tilde{\bm{F}}_{tea}^{(k,j)} =  \bm{F}_{tea}^{(k,j)} \cdot \bm{F}_{tea}^{{(k,j)}^T}; G_{{tea}_{[i,:]}}^{(k,j)}=\Tilde{\bm{F}}_{{tea}_{[i,:]}}^{(k,j)}/||\Tilde{\bm{F}}_{{tea}_{[i,:]}}^{(k,j)}||_2 \\
&\Tilde{\bm{F}}_{stu}^{(k,j)} =  \bm{F}_{stu}^{(k,j)} \cdot \bm{F}_{stu}^{{(k,j)}^T};
G_{{stu}_{[i,:]}}^{(k,j)}=\Tilde{\bm{F}}_{{stu}_{[i,:]}}^{(k,j)}/||\Tilde{\bm{F}}_{{stu}_{[i,:]}}^{(k,j)}||_2
\end{aligned}
\end{equation}
where $[i, :]$ denotes the $i$-th row in a matrix and the dimension of the similarity matrix $G_{[i,:]}^{(k,j)}$ for each frame is $b\times b$. The SPKD loss for the $k^{th}$ layer is then calculated as the sum of the similarity distances across all frames. The overall SPKD loss is determined by summing the SPKD loss of each layer as follows
\begin{equation}
\mathcal{L}_{KD}^{SPKD} = \sum_{k=1}^{K}\mathcal{L}_{SPKD}^k = \frac{1}{b^2}\sum_{k=1}^{K}\sum_{j=1}^{S'} ||\bm{G}_{tea}^{(k,j)}- \bm{G}_{stu}^{(k,j)}||_F^2,
\label{Loss: KD_SPKD}
\end{equation}
where $||\cdot||_F$ is the Frobenius norm.

Therefore, to train the proposed KD-based audio-lip SE model, the overall loss function can be written as
\begin{equation}
\mathcal{L}_{KD} = \mathcal{L}_{SE} + \delta_1\mathcal{L}_{KD}^{MSE} + \delta_2\mathcal{L}_{KD}^{SPKD},
\label{Loss: KD}
\end{equation}
where, $\delta_1$ and $\delta_2$ are all hyperparameters used to ensure that the loss functions are scaled to the similar magnitude.

The KD-based audio-lip SE method demonstrates the preliminary feasibility of utilizing tongue information even when only two modalities are employed during inference. It provides the advantage of eliminating the need for additional parameters compared to the audio-lip SE model, as it solely relies on supervision signals from the audio-lip-tongue teacher model during training to acquire certain aspects of the tongue knowledge embedded within the teacher model. 
Nevertheless, owing to the inherent nature of KD, where the teacher model often serves as the performance upper bound for the student model, a performance gap persists between the audio-lip SE model trained using KD and the audio-lip-tongue SE model. 
Additionally, since the KD-based method implicitly transfers the tongue information contained in the audio-lip-tongue SE model to the student model, it ignores the alignment between the tongue feature and the lip feature, which may also cause the suboptimal performance of the KD-based audio-lip SE model.

\subsection{Memory-Based Audio-Lip SE Method}
\label{Section: memory-based}

\begin{figure*}[t]
  \centering
  \includegraphics[width=0.99\linewidth]{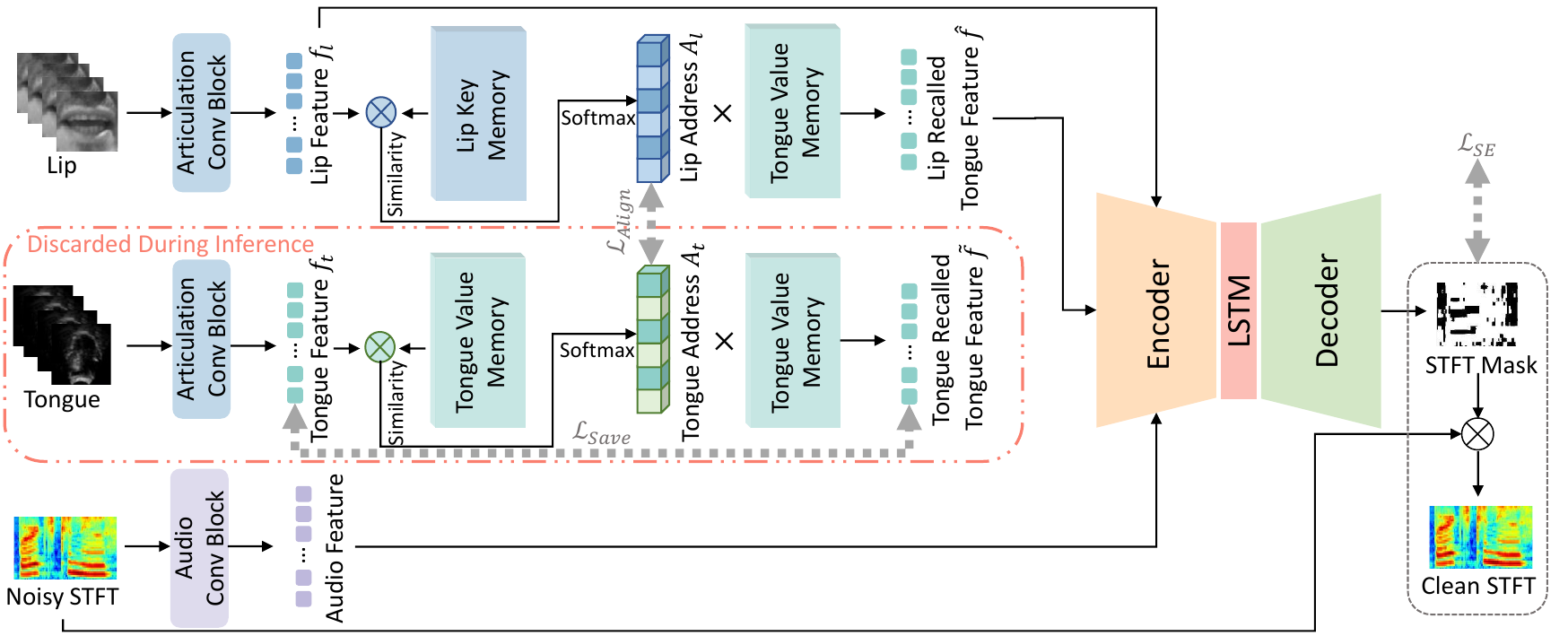}
  \caption{Details of the proposed memory-based audio-lip SE approach. The dashed arrows indicate the loss function. The lip-tongue memory is inserted right after the articulation convolutional block of the audio-lip-tongue SE encoder. The architecture of the model's rest part is identical with the audio-lip-tongue SE model depicted in Fig.\ref{Figure: Teacher and Student}(a). }
  \label{Figure: Memory}
\end{figure*}

In KD methods, the performance of the teacher model sets an upper limit which is often challenging for the student model to reach. Additionally, the KD-based audio-lip SE method implicitly fuses lip and tongue features while transferring tongue knowledge, making it challenging to learn the alignment relationship between these two modalities. To address these limitations, we propose a novel approach to better model the alignment between lip and tongue modalities and reach the upper limit set by the audio-lip-tongue SE model. Drawing inspiration from \cite{kim2021multi, park2022synctalkface}, we introduce a lip-tongue key-value memory network into the encoder of the audio-lip-tongue SE model, creating a memory-based audio-lip SE model. The architecture of this memory-based audio-lip SE method is illustrated in Figure \ref{Figure: Memory}.

The lip features $\bm{F}_{l}\in \mathbb{R}^{C\times S'\times D}$ and tongue features $\bm{F}_{t} \in \mathbb{R}^{C\times S'\times D}$ are obtained from the corresponding articulation convolutional block in the audio-lip-tongue SE encoder and then transformed through a simple reshaping operation into $\bm{F}_{l}$ $=[\bm{f}_{l}^1, \bm{f}_{l}^2, \dots, \bm{f}_{l}^{S'}]^T \in \mathbb{R}^{S'\times D’}$ and $\bm{F}_{t}$ $=[\bm{f}_{t}^1, \bm{f}_{t}^2, \dots, \bm{f}_{t}^{S'}]^T \in \mathbb{R}^{S'\times D’}$, where $D'=C\times D$ denotes each feature's dimension. The primary objective of this method is to store the tongue features $\bm{F}_{t}$ in the tongue value memory and recall them using lip features $\bm{F}_{l}$, which can be easily obtained from lip videos. 
To this end, two trainable modality-specific memories are introduced: the lip key memory $\bm{M}_{l}=[\bm{m}_{l}^1, \bm{m}_{l}^2, \dots, \bm{m}_{l}^N]^T \in \mathbb{R}^{N\times D'}$ and the tongue value memory $\bm{M}_{t}=[\bm{m}_{t}^1, \bm{m}_{t}^2, \dots, \bm{m}_{t}^N]^T \in \mathbb{R}^{N\times D'}$, where $N$ denotes the number of memory slots, and $\bm{m}_{l}^i, \bm{m}_{t}^i \in \mathbb{R}^{D'}$ correspond to the distinct lip and tongue features in the $i$-th slot, respectively. These memories are inserted right after the first articulation convolutional block of the encoder in the audio-lip-tongue SE model shown in Figure \ref{Figure: Teacher and Student}(a). The remaining architecture of the model illustrated in Figure \ref{Figure: Memory} remains identical to that of the audio-lip-tongue SE model depicted in Figure \ref{Figure: Teacher and Student}(a).

Each memory in the proposed method is a randomly initialized matrix at the beginning and is then trained to store representative lip and tongue features in a paired manner. 
During training, an associative alignment between the lip key memory $\bm{M}_{l}$ and the tongue value memory $\bm{M}_{t}$ is established, enabling the framework to access the tongue value memory $\bm{M}_{t}$ by querying the lip key memory $\bm{M}_{l}$ with the lip features $\bm{F}_{l}$. 
During inference, when lip features are provided as input, the stored tongue features $\bm{F}_{t}$ can be retrieved from the tongue value memory $\bm{M}_{t}$ through this associative alignment. 
This allows the model to complement the information of uni-modal lip videos with the recalled tongue features, thereby enhancing the capability of the model in solving the downstream AV-SE task. 
As a result, the memory-integrated audio-lip-tongue SE model can convert to an audio-lip SE model during inference, eliminating the need for ultrasound tongue inputs to obtain tongue features.
Further details of the proposed memory-based method are discussed next.

\subsubsection{Generating Addressing Vectors} 
We start by introducing how the source lip and target tongue addressing vectors are formulated. These addressing vectors are critical in determining the weights assigned to the memory slots for a given query. Expressly, when the lip key features $\bm{F}_{l}$ are provided as a query, the cosine similarity distance between each lip feature frames $\bm{f}_{l}^j$ and each lip key memory slots $\bm{m}_{l}^i$ is computed as follows:
\begin{equation}
    d_{l}^{i,j} = \frac{\bm{m}_{l}^i \cdot \bm{f}_{l}^j}{||\bm{m}_{l}^i||_2 \cdot ||\bm{f}_{l}^j||_2},
    \label{Equation: distance}
\end{equation}
where $d_{l}^{i,j}$ represents the cosine similarity distance between $i$-th memory slot of lip key memory and lip features in $j$-th temporal step. The relevance probability is then obtained using the Softmax function as follows,
\begin{equation}
    \alpha_{l}^{i,j} = \frac{\exp(\gamma\cdot d_{l}^{i,j})}{\sum_{i=1}^N \exp(\gamma\cdot d_{l}^{i,j})},
    \label{Equation: Addressing Vectors}
\end{equation}
where $\gamma$ is a scaling factor for similarity. It regulates the uniformity of the Softmax output distribution. A higher $\gamma$ value results in a more concentrated distribution, whereas a lower $\gamma$ value leads to a smoother distribution. By calculating the probability over the entire memory slot, the lip addressing vector for the $j$-th temporal step $\bm{a}_{l}^j = [\alpha_{l}^{1,j}, \alpha_{l}^{2,j}, \dots, \alpha_{l}^{N,j}]^T$ can be obtained.

The same procedure is applied to target tongue features $\bm{F}_{t}$ and tongue value memory $\bm{M}_{t}$ to generate the tongue addressing vector $\bm{a}_{t}^j = [\alpha_{t}^{1,j}, \alpha_{t}^{2,j}, \dots, \alpha_{t}^{N,j}]^T$ for the $j$-th temporal step. The addressing vector plays a crucial role in recalling the stored tongue features from the tongue value memory and establishing the alignment between the two modality-specific memories.

\subsubsection{Saving Tongue Features in Tongue Value Memory} 
The obtained tongue addressing vector $\bm{a}_{t}^j$ aims to accurately match the tongue value memory $\bm{M}_{t}$ in order to reconstruct the tongue features $\bm{F}_{t}$. The tongue value memory $\bm{M}_{t}$ is trained to save the appropriate tongue features $\bm{F}_{t}$ for retrieval. The $j$-th recalled tongue feature frame are obtained as follows:
\begin{equation}
    \bm{\Tilde{f}}_{t}^j = {\bm{M}_{t}}^T\bm{a}_{t}^j.
    \label{Equation: Retrieval}
\end{equation}
Next, a saving loss function is designed to guide the tongue value memory $\bm{M}_{t}$ to save the appropriate tongue features. This loss function aims to minimize the Euclidean distance between the reconstructed tongue features $\bm{\Tilde{f}}_{t}^j$ and the real tongue features $\bm{f}_{t}^j$. It can be expressed as follows,
\begin{equation}
    \mathcal{L}_{Save} = \mathbb{E}_j[||\bm{f}_{t}^j - \bm{\Tilde{f}}_{t}^j||_2^2].
    \label{Loss: Save}
\end{equation}
By using the saving loss, the tongue value memory $\bm{M}_{t}$ is trained to save the representative tongue features. As a result, the recalled tongue features $\bm{\Tilde{f}}_{t}^j$ retrieved from tongue value memory $\bm{M}_{t}$ are capable of representing the original tongue features $\bm{f}_{t}^j$ effectively.

\subsubsection{Aligning Lip Address with Tongue Address} 
To recall the tongue features $\bm{F}_{t}$ from the tongue value memory $\bm{M}_{t}$ using lip features $\bm{F}_{l}$ as inputs, an associative alignment between the lip key memory $\bm{M}_{l}$ and the tongue value memory $\bm{M}_{t}$ is constructed. Specifically, the lip key memory $\bm{M}_{l}$ is utilized to generate the lip addressing vector $\bm{a}_{l}^j$,  while the tongue addressing vector $\bm{a}_{t}^j$ is produced from the tongue value memory $\bm{M}_{t}$. The alignment between the two vectors is achieved by minimizing the Kullback-Leibler divergence with the alignment loss:
\begin{equation}
    \mathcal{L}_{Align} = \mathbb{E}_j[D_{KL}(\bm{a}^j_{t}||\bm{a}^j_{l})],
    \label{Loss: Align} 
\end{equation}
where $D_{KL}(\cdot)$ represents Kullback–Leibler divergence. By incorporating the alignment loss, the lip key memory $\bm{M}_{l}$ saves the lip features $\bm{F}_{l}$ in the same location, where the tongue value memory $\bm{M}_{t}$ stores the corresponding tongue features $\bm{F}_{t}$. Therefore, when a lip feature $\bm{F}_{l}$ is given, the lip key memory $\bm{M}_{l}$ provides the location information of the corresponding saved tongue features $\bm{F}_{t}$ in the tongue value memory $\bm{M}_{t}$, using the lip addressing vector $\bm{a}_l^j$.

\subsubsection{Applying Memory Network for AV-SE} Through the utilization of modality-specific memories and the associative alignment, the recalled tongue features $\bm{\hat{F}}_t$ can be acquired by multiplying the lip addressing vectors $\bm{a}_l^j$ with the tongue value memory $\bm{M}_{t}$:
\begin{equation}
    \bm{\hat{f}}_t^j = {\bm{M}_{t}}^T\bm{a}_l^j.
    \label{Equation: recalled by lip}
\end{equation}

In this process, the target tongue features $\bm{\hat{F}}_t = [\bm{\hat{f}}_t^1, \bm{\hat{f}}_t^2, \dots, \bm{\hat{f}}_t^{S'}]^T$ are recalled by querying the lip key memory $\bm{M}_{l}$ with the source lip features $\bm{F}_l$. Thus, there is no requirement for ultrasound tongue images as input to generate the tongue features anymore, and the memory-integrated audio-lip-tongue SE model can convert to an audio-lip SE model during inference. The recalled tongue features $\bm{\hat{F}}_t$ can then be utilized alongside the lip features $\bm{F}_l$ for the downstream AV-SE task, effectively enhancing task performance by leveraging the complementary information. During the training of the proposed memory-based method, the following loss function is employed:
\begin{equation}
    \mathcal{L}_{Mem} = \mathcal{L}_{SE} + \beta_1\mathcal{L}_{Save} + \beta_2\mathcal{L}_{Align}.
    \label{Loss: memory}
\end{equation}
The calculation method for $L_{SE}$ is the same as the definition in Eq.(\ref{Loss: SE}), with the difference that Eq.(\ref{Loss: SE}) uses tongue features obtained from the real ultrasound tongue images, while here tongue features retrieved by lip features are employed for the subsequent SE task.

The memory-based audio-lip SE model explicitly models the alignment relationship between lip and tongue features. It demonstrates remarkable performance that is nearly comparable to the audio-lip-tongue SE model, and even surpasses it in certain metrics. These results highlight the effectiveness of leveraging the modality-specific memories and associative alignment for enhancing the model's performance in the absence of ultrasound tongue images during inference.

\section{Implementation Details}
\label{Section: Implementation Details}
\subsection{Datasets}
\label{Section: Datasets}
For our main experiments, we utilized the TaL corpus \cite{ribeiro2021tal}, a multi-speaker dataset containing ultrasound tongue imaging, optical lip videos, and audio for each utterance. We focused on the TaL80 subset, which includes recordings from 81 native English speakers without any voice talent. Each utterance in TaL80 contains ultrasound tongue frames recorded at a frame rate of 81.5fps, a corresponding lip video recorded at 60fps, and audio recorded at a sampling frequency of 48kHz with 16-bit depth. In our experiments, the training set comprised 10,271 utterances from 73 speakers, while the validation set comprised 810 utterances from 81 speakers (including eight speakers not present in the training set). The test set consisted of 1,749 utterances from 73 speakers seen during training and 1,407 utterances from eight unseen speakers. The content of the three sets was mutually exclusive from each other.

To further demonstrate the effectiveness of incorporating ultrasound tongue images in our proposed methods and assess their generalization potential for the AV-SE task, additional experiments were conducted  using a subset of the GRID dataset \cite{cooke2006audio}, a commonly used database in the area of deep-learning-based AV-SE \cite{michelsanti2021overview}. For these experiments, 12 speakers were picked up from the GRID dataset \cite{cooke2006audio}. In the training dataset, 8,980 utterances from 10 speakers were randomly selected , while 600 utterances from all 12 speakers were used for the validation and test sets respectively. The content of the three sets was mutually exclusive from each other.

The noisy speech was generated by mixing noise with the clean speech. During training, we introduced ten different noise types as in \cite{valentini2018speech}: eight noise recordings from the DEMAND database \cite{thiemann2013diverse} and two artificially generated noises, namely speech-shaped noise and babble noise. These noise types were added to the speech signal at three signal-to-noise (SNR)  values (0dB, -5dB, and -10dB). For validation and testing, we added five additional noises: living room, office, bus, street cafe, and a public square. We used slightly higher SNR values (2.5dB, -2.5dB, and -7.5dB) than the ones used during training following the previous work \cite{valentini2018speech}.

\subsection{Experimental Setups}
\label{Section: Experimental Setups}
The lip videos and ultrasound tongue images in the TaL80 dataset \cite{ribeiro2021tal} were preprocessed following the pipeline described in \cite{zhang2021talnet}. The resulting lip videos and ultrasound tongue images were resized to $64\times128$ and had a frame rate of 81.5 fps. During the preprocessing of the GRID \cite{cooke2006audio} subset, the lip videos were initially upsampled from 25fps to 81.5 fps. Subsequently, the mouth Regions of Interest (ROIs) were cropped to a size of $64 \times 128$, consistent with the settings used in preprocessing the TaL80 dataset. For audio preprocessing, the audio was downsampled to 16 kHz, and the STFT was computed using a Hann window with a length of 512, a hop size of 196, and an FFT point number of 512 to match the frame rate of the ultrasound. The real and imaginary parts of the spectrogram were concatenated along the channel axis, resulting in a complex spectrogram with dimensions of $2\times257\times T$, where $T$ represents the number of frames.

All SE models in this study were implemented using the PyTorch library \cite{paszke2019pytorch}. The Adam optimizer\cite{kingma2014adam} with an initial learning rate of 1e-3 was used for all experiments. The learning rate was decreased by a factor of 0.1 once learning stagnated, i.e., the validation error did not improve for ten epochs. The models were trained at the utterance level by randomly cropping all the samples in the mini-batch to have the same number of frames as the shortest one. The training was conducted with a batch size of 8 on a single NVIDIA GeForce GTX 3090 for 150 epochs. Cross-validation was employed to select the best-performing model for testing.

During the training of the audio-lip-tongue SE model and the KD-based audio-lip SE model, the hyperparameters $\alpha$ in Eq. (\ref{Loss: SE}) and $\delta_1, \delta_2$ in Eq. (\ref{Loss: KD}) were empirically set to ensure comparable scaling of the multiple loss functions. Before training the memory-based audio-lip SE model, the weights of the corresponding part from a pre-trained audio-lip-tongue SE model were loaded into the proposed memory-based audio-lip SE model. For training the memory-based audio-lip SE model, the hyperparameters $\beta_1$, $\beta_2$ in Eq.(\ref{Loss: memory}) and $\gamma$ in Eq.(\ref{Equation: Addressing Vectors}) were empirically set to 0.01, 0.001 and 1, respectively. The number of slots $N$ in the memory was set to 512 through hyperparameter selection experiment presented in Section \ref{Section: mem-ablation}.

While evaluating the generalization potential of our proposed methods on the GRID subset, each of the proposed audio-lip, KD-based, and memory-based models, originally trained on the TaL80 dataset, were fine-tuned on the modified GRID subset with a learning rate of 5e-5. It's crucial to note that GRID datasets lack ultrasound tongue images, and consequently, each model was trained exclusively using audio and lip videos. In other words, during the fine-tuning process, each system was trained solely with the loss function $\mathcal{L}_{SE}$ defined in Eq.(\ref{Loss: SE}). Furthermore, the memory network in the memory-based SE model was kept frozen, functioning to retrieve tongue features.

\subsection{Evaluation Metrics}
\label{Section: Evaluation metrics}
Several evaluation metrics were employed to assess the performance of our proposed methods, providing quantitative measures of the quality and intelligibility of the enhanced speech from different dimensions. These include Segmental Signal-to-Noise Ratio (SegSNR) for energy-based speech quality assessment, Perceptual Evaluation of Speech Quality (PESQ) for speech quality assessment based on perceptual models, and Short-Time Objective Intelligibility (STOI) for measuring speech intelligibility. These metrics offer comprehensive insights into the effectiveness of our AV-SE methods.

\subsection{Baselines}
\label{Section: Baselines}
A commonly used open-source AO-SE baseline model named DEMUCS \cite{defossez2020real} was adopted as our AO-SE baseline model. It is built upon a causal architecture that utilizes convolutions and LSTMs. It operates on waveform domain through a hierarchical generation process, employing U-Net-like skip connections for enhanced performance.

Four recent lip-based AV-SE methods \cite{gabbay2018visual, hegde2021visual, hussain2021towards, gao2021visualvoice} whose source codes were available online were used as baselines for comparison with our proposed audio-lip SE models. 
\subsubsection{VSE} The VSE model \cite{gabbay2018visual} takes video frames of the speaker's mouth and a spectrogram of the noisy audio as inputs. It employs an encoder-decoder structure, where the encoder module, comprising a dual tower CNN, encodes the audio-visual features into a shared embedding. The decoder module, consisting of transposed convolutional layers, decodes the shared embedding into a spectrogram of the enhanced speech. 
\subsubsection{PVSE} The PVSE model \cite{hegde2021visual} utilizes both pseudo visual and noisy auditory inputs. It includes a visual encoder and a speech encoder that process lip movements and noisy spectrograms. The speech decoder produces a residual mask that is applied to the input spectrograms to filter out noise from the clean speech. It is worth noting that when evaluating PVSE, as a generated pseudo lip stream was adopted in its original paper, we used the natural lip stream for a fair comparison. 
\subsubsection{IOAVSE} The IOAVSE model \cite{hussain2021towards} focuses on addressing the challenges of AV-SE by introducing intelligibility-oriented (I-O) loss functions. This is achieved by involving a deep learning-based fully convolutional AV-SE model and utilizing a modified short-time objective intelligibility (STOI) metric as the training cost function. 
\subsubsection{VisualVoice} The VisualVoice model \cite{gao2021visualvoice} presents a multi-task learning framework designed to simultaneously solve AV-SE task using facial appearance, lip motion, and vocal audio, while also learning cross-modal speaker embeddings. During the training phase, this model incorporates not only the traditional mix-and-separate loss but also introduces a cross-modal matching loss and a speaker consistency loss. However, it's important to note that when using VisualVoice as our baseline model, limitations were encountered due to the absence of face images in the TaL80 dataset and consequently certain adjustments were made. Specifically, face attribute analysis network from the VisualVoice model were omitted and only lip motion was used as the visual feature. In this modified setup, our objective function consisted of the mix-and-separate loss and the speaker consistency loss.

Each baseline was trained with the codes available online on our constructed dataset, following the training strategy in its original paper.

\section{Experimental Results}
\label{Section: Experimental Results}

\subsection{Overall Performance}
\label{Section: Overall Performance}

\begin{table*}[htbp]
\caption{Evaluation results of the proposed methods compared with baseline models. The first column denotes the modalities entered into the system during inference. Best results using noisy speech and lip videos as input during inference are highlighted in \textbf{bold} and \uline{Underline} characters indicate the sub-optimal results.}
\centering
\label{Table: Overall Performance}
\begin{tabular}{c|cc|ccc|ccc|ccc}
\toprule
\toprule
\multirow{2}{*}{\begin{tabular}[c]{@{}c@{}}Inference \\ Modality\end{tabular}} & \multicolumn{2}{c|}{\multirow{2}{*}{Method}}     & \multicolumn{3}{c|}{SNR=2.5dB}                          & \multicolumn{3}{c|}{SNR=-2.5dB}                         & \multicolumn{3}{c}{SNR=-7.5dB}                          \\ \cline{4-12} 
                                                                               & \multicolumn{2}{c|}{}                        & SegSNR          & PESQ            & STOI            & SegSNR          & PESQ            & STOI            & SegSNR          & PESQ            & STOI            \\ \hline
/                                                                              & \multicolumn{2}{c|}{Noisy}                             & 0.2414          & 1.4285          & 0.8503          & -3.1703         & 1.2283          & 0.7527          & -5.9847         & 1.1298          & 0.6229          \\ \hline
\multirow{6}{*}{Audio+Lip}                                                     & \multicolumn{2}{c|}{VSE \cite{gabbay2018visual}}                               & 4.8731          & 1.8260          & 0.8871          & 3.8721          & 1.6502          & 0.8566          & 2.6978          & 1.4590          & 0.8005          \\
                                                                               & \multicolumn{2}{c|}{PVSE \cite{hegde2021visual}}                              & 3.6086          & 1.8756          & 0.8882          & 2.9790          & 1.6890          & 0.8603          & 2.0987          & 1.4863          & 0.8052          \\
                                                                               & \multicolumn{2}{c|}{IOAVSE \cite{hussain2021towards}}                            & 6.1654          & 2.0388          & 0.9245          & 4.5055          & 1.7750          & 0.8867          & 2.8217          & 1.5424          & 0.8177          \\
                                                                                & \multicolumn{2}{c|}{VisualVoice \cite{gao2021visualvoice}}                            & 7.5556          & 2.0377          & 0.9006          & 6.0683          & 1.7899          & 0.8735          & 4.3432          & 1.5643          & \textbf{0.8278}         \\
                                                                               & \multicolumn{2}{c|}{Audio-Lip}                 & 8.9741          & 2.0841          & 0.9304          & 6.8073          & 1.8030          & 0.8854          & 4.6314          & 1.5604          & 0.8142          \\
                                                                               & \multicolumn{2}{c|}{KD-Based}                          & {\ul 9.0061}    & {\ul 2.1073}    & {\ul 0.9355}    & {\ul 6.8343}    & {\ul 1.8241}    & {\ul 0.8911}    & {\ul 4.6944}    & {\ul 1.5858}    & 0.8221    \\
                                                                               & \multicolumn{2}{c|}{Memory-Based}                      & \textbf{9.2291} & \textbf{2.1442} & \textbf{0.9356} & \textbf{7.0638} & \textbf{1.8548} & \textbf{0.8956} & \textbf{4.8593} & \textbf{1.6020} & \uline{0.8264} \\ \hline
Audio                                                                          & \multicolumn{2}{c|}{DEMUCS \cite{defossez2020real}}                        & 8.1642          & 1.7190          & 0.9215          & 5.9709          & 1.4794          & 0.8809          & 3.9411          & 1.2960          & 0.8090          \\

                                                                                & \multicolumn{2}{c|}{Audio-Only}                        & 8.6748          & 1.9872          & 0.9215          & 6.4224          & 1.7088          & 0.8661          & 4.0888          & 1.4773          & 0.7739          \\ \hline
Audio+Tongue                                                                   & \multicolumn{2}{c|}{Audio-Tongue}                      & 8.8509          & 2.0990          & 0.9321          & 6.7223          & 1.8133          & 0.8901          & 4.6153          & 1.5631          & 0.8220          \\ \hline
Audio+Lip+Tongue                                                               & \multicolumn{2}{c|}{Audio-Lip-Tongue}                  & 9.0315          & 2.1122          & 0.9370          & 6.9267          & 1.8322          & 0.8987          & 4.8707          & 1.5973          & 0.8394         \\
\bottomrule
\bottomrule
\end{tabular}
\end{table*}

The SE performance of the two proposed audio-lip SE methods at different SNRs is provided in Table \ref{Table: Overall Performance}. Additionally, for reference, we included results from the well-trained audio-lip-tongue SE model together with the audio-tongue and audio-only SE models, which were built by replacing lip inputs with tongue inputs or removing the lip inputs of the proposed audio-lip SE model depicted in Fig.\ref{Figure: Teacher and Student}(b). The results of the proposed audio-lip SE model trained solely with clean speech supervision were also included. \footnote{Speech samples are available at \url{https://zhengrachel.github.io/IUTIforAVSE-demo/}}

By comparing the results of the AO-SE baseline with the lip-based AV-SE baselines, it is evident that the lip-based AV-SE baselines exhibited superior performance on the PESQ metric, particularly at low Signal-to-Noise Ratios (SNRs). The advantage of incorporating lip information into SE task becomes more obvious when comparing the DEMUCS \cite{defossez2020real} and VisualVoice \cite{gao2021visualvoice} models. The lip-based AV-SE baselines \cite{gabbay2018visual, hegde2021visual, hussain2021towards} showed poorer performance in terms of SegSNR, possibly due to these models mainly operating in the magnitude domain without accounting for phase information. Moreover, the results clearly demonstrate that both of our proposed methods outperformed all the baselines and the audio-lip SE student model across most evaluation metrics when audio and lip videos were used as input modalities during inference. The only exception was a slightly inferior STOI performance when SNR was -7.5 dB compared to VisualVoice \cite{gao2021visualvoice}. These findings suggest a significant improvement in speech quality and intelligibility and highlight the effectiveness of incorporating tongue information through the two proposed methods. Furthermore, the performance of the two proposed methods outperformed that of all the other modalities input models except the audio-lip-tongue SE model, emphasizing the effectiveness of combining multi-modal sources of articulation knowledge. 

It is also worth noting that the memory-based audio-lip SE model surpasses the KD-based audio-lip model on all metrics, and meets or even exceeds the performance of the audio-lip-tongue SE model on most metrics, indicating its superior capability to incorporate tongue information into the lip-based AV-SE system. 

\subsection{Generalization to Unseen Speakers and Noises}
\label{Section: generalization}

\begin{table*}[]
\caption{Generalization performance of the proposed methods. ``Seen Noise" and ``Unseen Noise" here distinguish the noise types. The numbers in (parentheses) indicate the number of test utterances of this setting. Best results are highlighted in \textbf{bold}. \uline{Underline} characters indicate the sub-optimal results.}
\label{Table: Generalization}
\centering
\begin{tabular}{cc|ccc|ccc|ccc|ccc}
\toprule
\toprule
\multicolumn{2}{c|}{\multirow{2}{*}{Method}} & \multicolumn{3}{c|}{\begin{tabular}[c]{@{}c@{}}Seen Speakers \\ \& Seen Noise (1160)\end{tabular}} & \multicolumn{3}{c|}{\begin{tabular}[c]{@{}c@{}}Seen Speakers \\\& Unseen Noise (589)\end{tabular}} & \multicolumn{3}{c|}{\begin{tabular}[c]{@{}c@{}}Unseen Speakers \\ \& Seen Noise (900)\end{tabular}} & \multicolumn{3}{c}{\begin{tabular}[c]{@{}c@{}}Unseen Speakers \\ \& Unseen Noise (507)\end{tabular}} \\ \cline{3-14} 
\multicolumn{2}{c|}{}                        & SegSNR                            & PESQ                              & STOI                             & SegSNR                            & PESQ                              & STOI                              & SegSNR                            & PESQ                              & STOI                              & SegSNR                             & PESQ                              & STOI                              \\ \hline
\multicolumn{2}{c|}{Noisy}                       & -2.9905                           & 1.1826                            & 0.6767                           & -2.6012                           & 1.3687                            & 0.8130                            & -3.0473                            & 1.2144                            & 0.7275                           & -2.8073                             & 1.4286                            & 0.8470                            \\ \hline
\multicolumn{2}{c|}{VSE\cite{gabbay2018visual}}                       & 4.2344                            & 1.6151                            & 0.8288                           & 4.9624                            & 1.7890                            & 0.8661                            & 2.8825                            & 1.5693                            & 0.8478                            & 3.3009                             & 1.7030                            & 0.8760                            \\
\multicolumn{2}{c|}{PVSE\cite{hegde2021visual}}                      & 3.2984                            & 1.6592                            & 0.8378                           & 3.3542                            & 1.7723                            & 0.8554                            & 2.4558                            & 1.6465                            & 0.8577                            & 2.3031                             & 1.7259                            & 0.8696                            \\
\multicolumn{2}{c|}{IOVSE\cite{hussain2021towards}}                     & 4.7901                            & 1.6943                            & 0.8529                           & 5.9487                            & 1.8609                            & 0.8943                            & 3.4277                            & 1.7705                            & 0.8848                            & 4.2480                             & \uline{1.9650}                            & 0.9198                            \\ 
\multicolumn{2}{c|}{VisualVoice\cite{gao2021visualvoice}}                     & \uline{6.7832}                            & \textbf{1.8836}                            & \textbf{0.8742}                           & \uline{8.5968}                            & \textbf{2.1071}                            & \textbf{0.9215}                            & 3.9189                            & 1.5641                            & 0.8288                            & 5.0057                             & 1.6842                            & 0.8605                            \\ \hline
\multicolumn{2}{c|}{KD-Based}                                 & 6.7830                            & 1.7684                            & 0.8534                           & 8.5416                           & 1.9434                           & 0.9058                            & \uline{5.7546}                            & \uline{1.8147}                            & \uline{0.8835}                            & \uline{7.2204}                             & 1.9573                            & \uline{0.9239}                            \\
\multicolumn{2}{c|}{Memory-Based}                & \textbf{6.9916}                            & \uline{1.7947}                           & \uline{0.8578}                           & \textbf{8.6977 }                           & \uline{1.9815 }                           & \uline{0.9082}                            & \textbf{5.9689}                            & \textbf{1.8385 }                           & \textbf{0.8874}                            & \textbf{7.4629 }                            & \textbf{1.9857}                            & \textbf{0.9269}                         \\
\bottomrule
\bottomrule
\end{tabular}
\end{table*}

The generalization capability of the models is a significant concern in AV-SE, as it is closely related to real-world applications where the SE system encounters diverse and unpredictable acoustic conditions. To assess the performance of the proposed methods in diverse scenarios, we further partitioned the test set into four distinct categories: seen speakers with seen noises types, seen speakers with unseen noises types, unseen speakers with seen noises types, and unseen speakers with unseen noises types. The evaluation results for these scenarios are presented in Table \ref{Table: Generalization}. The results clearly indicate that though VisualVoice \cite{gao2021visualvoice} achieved the best performance in the context of seen speakers, it falls short in terms of generalizing effectively to unseen speakers. In contrast, both of our proposed methods outperform all the baselines even in the challenging scenario of unseen speakers with unseen noise types. This underscores the robust generalization capability of our methods and suggests their potential for practical deployment in AV-SE applications. One point to clarify is that the performance discrepancy between the seen and unseen noise types arises due to the specific challenges posed by the babble noise contained in the seen noise types, which is generated by mixing the utterances of multiple speakers and harder to perform noise reduction than other noise types.

\subsection{Generalization to Other Datasets}

\begin{table*}[]
\centering
\caption{Evaluation results of the audio-lip and the two proposed methods on the constructed GRID datasets. Best results are highlighted in \textbf{bold} and \uline{Underline} characters indicate the sub-optimal results. Audio-Lip* represents the evaluation results of the audio-lip model trained on the TaL80 dataset without additional fine-tuning on the GRID dataset.}
\label{Table: GRID}
\begin{tabular}{cl|ccc|ccc|ccc}
\toprule
\toprule
\multicolumn{2}{c|}{\multirow{2}{*}{Method}} & \multicolumn{3}{c|}{2.5dB}                          & \multicolumn{3}{c|}{-2.5dB}                         & \multicolumn{3}{c}{-7.5dB}                          \\ \cline{3-11} 
\multicolumn{2}{c|}{}                        & SegSNR          & PESQ            & STOI            & SegSNR          & PESQ            & STOI            & SegSNR          & PESQ            & STOI            \\ \hline
\multicolumn{2}{c|}{Noisy}                   & -2.3596         & 1.4587          & 0.7948          & -4.8332         & 1.2288          & 0.7393          & -7.1352         & 1.1363          & 0.6493          \\ \hline
\multicolumn{2}{c|}{Audio-Lip*}               & 3.4044          & 1.7877          & 0.7684          & 2.0261          & 1.6049          & 0.7130          & 0.4396          & 1.4126          & 0.6326          \\
\multicolumn{2}{c|}{Audio-Lip}               & 6.0618          & 1.8950          & 0.8545          & 4.5573          & 1.7063          & 0.8163          & 2.7979          & 1.4943          & 0.7363          \\
\multicolumn{2}{c|}{KD-Based}                & \uline{6.1513}          & \textbf{1.9268} & \textbf{0.8569} & \uline{4.6734}          & \textbf{1.7414} & \textbf{0.8210} & \textbf{2.8569} & \textbf{1.5228} & \textbf{0.7433} \\
\multicolumn{2}{c|}{Memory-Based}            & \textbf{6.2120} & \uline{1.9211}          & \uline{0.8558}          & \textbf{4.7031} & \uline{1.7372}          & \uline{0.8205}          & \uline{2.8383}          & \uline{1.5159}          & \uline{0.7408}   \\
\bottomrule
\bottomrule
\end{tabular}
\end{table*}

To assess the generalization potential of our proposed methods to other datasets, we conducted additional experiments on the GRID \cite{cooke2006audio} subset. The experimental results are presented in Table \ref{Table: GRID}. For comparison, we additionally present the results from an audio-lip SE model without fine-tuning.

The comparison between the first and second rows of Table \ref{Table: GRID} indicates that any system needs fine-tuning to better adapt to new datasets. Moreover, it is evident that both of our proposed methods outperformed the audio-lip AV-SE method after fine-tuning, suggesting their potential for generalization to other datasets. The KD-based method exhibited slightly better speech enhancement performance compared to the memory-based method. 
This variance could possibly be attributed to the holistic updating of all parameters in the KD-based model during fine-tuning. In contrast, the parameters of the memory network in the memory-based model remained frozen due to the absence of ultrasound tongue images in the GRID subset. As a result, the tongue features restored using the memory-based method might lack generalization, possibly preserving certain dataset-dependent attributes.
Nonetheless, the memory-based method still remains more promising for generalization to other datasets when compared to purely lip-based AV-SE systems. The results of this analytical experiment serve as additional evidence of the effectiveness of incorporating ultrasound tongue images into the AV-SE task. Furthermore, these results illustrate that the generalization of our proposed methods to other datasets does not necessitate the collection of additional ultrasound tongue images, thereby reinforcing their potential applicability in real-world scenarios.

\subsection{Ablation and Hyperparameter Selection Experiments}
\label{Section: ablation}

\begin{table*}[]
\caption{Evaluation results of the proposed KD-based audio-lip SE model in ablation studies. Best results are highlighted in \textbf{bold}.}
\label{Table: Ablation KD}
\centering
\begin{tabular}{cc|ccc|ccc|ccc}
\toprule
\toprule
\multicolumn{2}{c|}{\multirow{2}{*}{Method}}     & \multicolumn{3}{c|}{SNR=2.5dB}                                                        & \multicolumn{3}{c|}{SNR=-2.5dB}                                                       & \multicolumn{3}{c}{SNR=-7.5dB}                                                       \\ \cline{3-11} 
\multicolumn{2}{c|}{}                       & \multicolumn{1}{c}{SegSNR} & \multicolumn{1}{c}{PESQ} & \multicolumn{1}{c|}{STOI} & \multicolumn{1}{c}{SegSNR} & \multicolumn{1}{c}{PESQ} & \multicolumn{1}{c|}{STOI} & \multicolumn{1}{c}{SegSNR} & \multicolumn{1}{c}{PESQ} & \multicolumn{1}{c}{STOI} \\ \hline
\multicolumn{2}{c|}{KD-Based}                         & \textbf{9.0061}               & 2.1073                   & \textbf{0.9335}              & \textbf{6.8343}               & \textbf{1.8241}             & \textbf{0.8911}              & \textbf{4.6944}               & \textbf{1.5858}             & \textbf{0.8221}             \\
\multicolumn{2}{c|}{w/o SPKD}                         & 8.9927                     & \textbf{2.1133}          & 0.9328                    & 6.8251                     & 1.8218                   & 0.8892                    & 4.6442                     & 1.5802                   & 0.8194                   \\
\multicolumn{2}{c|}{w/o KD (Audio-Lip)}                           & 8.9741                     & 2.0841                   & 0.9304                    & 6.8073                     & 1.803                    & 0.8854                    & 4.6314                     & 1.5604                   & 0.8142                   \\ 
\bottomrule
\bottomrule
\end{tabular}
\end{table*}

\begin{table*}[]
\caption{Evaluation results of the proposed memory-based audio-lip SE model with different number of memory slots. Best results are highlighted in \textbf{bold}. \uline{Underline} characters indicate the sub-optimal results.}
\label{Table: Ablation Mem}
\centering
\begin{tabular}{cc|ccc|ccc|ccc}
\toprule
\toprule
\multicolumn{2}{c|}{} & \multicolumn{3}{c|}{SNR=2.5dB} & \multicolumn{3}{c|}{SNR=-2.5dB} & \multicolumn{3}{c}{SNR=-7.5dB}  \\ \cline{3-11} 
\multicolumn{2}{c|}{\multirow{-2}{*}{Number of Slots}} & SegSNR & PESQ & \multicolumn{1}{c|}{STOI} & SegSNR & PESQ & \multicolumn{1}{c|}{STOI} & SegSNR & PESQ & STOI \\ \hline
\multicolumn{2}{c|}{128}                                                                 & 8.9949                         & 2.0907                       & 0.9277                                            & 6.9534                         & 1.7693                       & 0.8887                                            & 4.9481                         & 1.5722                       & 0.8141                       \\
\multicolumn{2}{c|}{256}                                         & 9.0217                         & 2.0957                       & 0.9295                                            & 6.9692                         & 1.7676                       & 0.8895                                            & 4.9311                         & 1.5706                       & 0.8130                       \\
\multicolumn{2}{c|}{512}                                                                 & \textbf{9.0755}                        & \textbf{2.1156 }                      & \uline{0.9284}                                            & \textbf{7.0311}                         & \textbf{1.7889}                       & \uline{0.8905}                                            & \uline{5.0076}                         & \uline{1.5830}                       & \uline{0.8157}                       \\
\multicolumn{2}{c|}{1024}                                        & 9.0516                         & 2.1034                       & \textbf{0.9294}                                           & 7.0228                         & 1.7859                       & \textbf{0.8912 }                                           & \textbf{5.0290}                         & \textbf{1.5882}                       & \textbf{0.8175}                      \\
\multicolumn{2}{c|}{2048}                                        & 8.9762                         & 2.0786                       & 0.9271                                            & 6.9365                         & 1.7610                       & 0.8873                                            & 4.9085                         & 1.5588                       & 0.8110                       \\
\multicolumn{2}{c|}{4096}                                        & 9.0724                         & 2.0783                       & 0.9259                                            & 6.9028                         & 1.7625                       & 0.8863                                            & 4.8828                         & 1.5616                       & 0.8099         \\
\bottomrule
\bottomrule
\end{tabular}
\end{table*}

\subsubsection{Ablation Studies for KD-based audio-lip SE Method}
\label{Section: KD-ablation}
Ablation study was conducted to provide further evidence of the efficacy of incorporating different KD loss in the KD-based audio-lip SE model. The outcomes are presented in Table \ref{Table: Ablation KD}. Specifically, the effectiveness of the SPKD loss described in Eq.(\ref{Loss: KD_SPKD}) was demonstrated by removing the loss function $\mathcal{L}_{KD}^{SPKD}$ in Eq.(\ref{Loss: KD}) (\textit{``w/o SPKD"}). To demonstrate the effectiveness of MSE Loss described in Eq.(\ref{Loss: KD_MSE}), an audio-lip SE model was trained solely with $\mathcal{L}_{SE}$ in Eq.(\ref{Loss: SE}) (\textit{``w/o KD (Audio-Lip)"}), which is exactly the same as that in Table \ref{Table: Overall Performance}. 

From the results, it can be concluded that the MSE loss effectively incorporates a portion of the articulation knowledge derived from the ultrasound tongue, since the outcomes of the \textit{``w/o SPKD"} surpassed those of the \textit{``w/o KD (Audio-Lip)"}.
Moreover, \textit{``w/o SPKD"} achieved results inferior to the proposed KD-based audio-lip SE method except for SNR=2.5dB, indicating the effectiveness of further introducing the SPKD loss particularly in low SNR scenarios. %

\subsubsection{Hyperparameter Selection for Memory-based Method}
\label{Section: mem-ablation}
In the memory network, the number of slots in the memory serves as a crucial hyperparameter. To determine the optimal value for it, we conducted experiments using different numbers of slots: 128, 256, 512, 1024, 2048, and 4096. The results 
on the validation set for these different parameter settings are presented in Table \ref{Table: Ablation Mem}. 

It can be observed that setting the number of slots to either 512 or 1024 yields the comparable best results. However, considering that the model's parameter size increases by approximately 10M when using 1024 slots compared to 512 slots, we chose 512 slots for all the experiments. Furthermore, Table \ref{Table: Ablation Mem} demonstrates a gradual improvement in the model's performance as the number of slots increases from 128 to 512/1024. This observation aligns with our intuition, as more slots can typically store more detailed information. However, unexpectedly, when the number of slots increases from 1024 to 4096, the model's performance starts to decline. This phenomenon could be attributed to the increased difficulty in aligning the lip key memory and tongue value memory during training as the number of slots increases. Therefore, as the number of slots further increases from 1024 to 4096, the loss function $\mathcal{L}_{Align}$ represented by Eq.(\ref{Loss: Align}) becomes increasingly challenging to minimize.

\subsection{Phoneme Error Rate Analysis}
\label{Section: PER}
To further study the gains of incorporating ultrasound tongue images, we employed an ASR engine to analyze the enhanced speech's PERs of different phoneme categories. An ASR API provided in ESPNet\footnote{\url{https://github.com/espnet/espnet_model_zoo}} \cite{watanabe2018espnet} was utilized to transcribe the enhanced speech. Each enhanced utterance was forcibly aligned with its transcription by the Montreal Forced Aligner (MFA)\footnote{\url{https://github.com/MontrealCorpusTools/Montreal-Forced-Aligner}} \cite{mcauliffe2017montreal} tool, consequently obtaining a time-aligned phoneme sequence. Ground truth phoneme sequences were obtained by aligning clean utterances with their corresponding texts using the MFA tool. The recognized and ground truth phoneme sequences were further aligned by minimizing the edit distance, and the PERs for different phoneme categories could be calculated. Phonemes were categorized according to the English (UK) MFA dictionary v2\_0\_0\footnote{\url{https://mfa-models.readthedocs.io/en/latest/dictionary/English/}}. 

\begin{table*}[]
\centering
\caption{PERs (\%) of different phoneme categories for clean speech, noisy speech, and the speech enhanced by different methods. Consonant phonemes are divided according to the place of articulation. Best results are highlighted in \textbf{bold}. \uline{Underline} characters indicate the sub-optimal results. The numbers in parentheses show relative PER reduction (\%) compared with the audio-only method.}
\label{Table: PER}
\begin{tabular}{c|c|c|cccccccc}
\toprule
\toprule
\multirow{2}{*}{Methods} & \multirow{2}{*}{Silence} & \multirow{2}{*}{Vowels} & \multicolumn{8}{c}{Consonants}                                                                                                                                                                                           \\ \cline{4-11} 
                         &                          &                         & Labial         & \begin{tabular}[c]{@{}c@{}}Labio-\\ dental\end{tabular} & Dental         & Alveolar       & \begin{tabular}[c]{@{}c@{}}Alveo-\\ palatal\end{tabular} & Palatal        & Velar          & Glottal        \\ \hline
Clean                    & 0.58                     & 3.09                    & 2.54           & 1.24                                                    & 2.18           & 2.96           & 1.94                                                     & 1.88           & 2.69           & 2.46           \\ 
Noisy                    & 25.10                    & 46.10                   & 41.58          & 37.44                                                   & 43.43          & 41.74          & 33.03                                                    & 37.53          & 37.10          & 46.35          \\ \hline
Audio-Only               & 10.54                    & 31.97                   & 33.54          & 34.03                                                   & 31.02          & 33.46          & 14.8                                                    & 30.78          & 27.37          & 33.76          \\  \hline
Audio-Lip                & 7.71                     & 27.67                   & 27.97          & 31.82                                                   & 24.78    & 30.29          & 12.48                                               & 30.33          & 26.48          & 30.42          \\
 & (26.85)                   & (13.45)                  & (16.61)         & (6.49)                                                  & (20.12)         & (9.47)          & (15.68)                                                   & (1.46)          & (3.25)          & (9.89)          \\   \hline
Audio-Lip-Tongue         & {\ul 6.34}            & \textbf{20.22}          & {\ul 24.95} & {32.15}                                          & {\ul 22.40} & \textbf{25.61} & {\ul 11.13}                                            & {\ul23.55} & {\ul17.35} & {\ul 25.09} \\
 & (39.84)                   & (36.75)                  & (25.61)         & (5.52)                                                   & (27.79)         & (23.46)         & (24.80)                                                   & (23.49)         & (36.61)         & (25.68)         \\  \hline
KD-Based                 & 7.34               &  25.31             &  26.42    &  {\ul 30.56}                                             & 25.44          &  29.35    &  11.17                                                     &  26.59    &  22.85    &  29.87    \\
 & (30.36)                   & (20.83)                  & (21.23)         & (10.20)                                                  & (17.99)         & (12.28)         & (24.53)                                                   & (13.61)         & (16.51)         & (11.52)        \\ \hline
 Memory-Based                & \textbf{6.18}               &  {\ul 22.84}             &  \textbf{21.93}    & \textbf{29.67}                                            & \textbf{22.19}          &  {\ul 25.95}    &  \textbf{10.55}                                                     &  \textbf{16.95}   &  \textbf{16.87}    &  \textbf{24.79}    \\
 & (41.37)                   & (28.56)                  & (34.62)         & (12.81)                                                  & (28.47)         & (22.44)         & (28.72)                                                   & (44.93)         & (38.36)         & (26.57)        \\ 
\bottomrule
\bottomrule  
\end{tabular}
\end{table*}

The results are shown in Table \ref{Table: PER}. Comparing audio-lip and audio-only methods, we can see that after introducing lip information for SE, PERs reduced significantly for some lip-related phonemes, such as silence, labials and dentals. However, for palatal and velar consonants, the relative PER reductions were quite small which indicates the limitations of using  only lip videos as articulation information. After further incorporating ultrasound tongue images, the audio-lip-tongue method achieved much more uniform PER reductions among phoneme categories than the audio-lip method. For palatals, velars and vowels whose articulations were mainly determined by internal tongue movement, their relative PER reductions improved significantly from 1.46\% to 23.49\%, from 3.25\% to 36.61\% and from 13.45\% to 36.75\%, respectively. Although not using tongue images at inference time, our proposed KD-based audio-lip SE method also achieved lower PERs than the audio-lip method for most phoneme categories, especially for palatals, velars and vowels, indicating the effectiveness of incorporating tongue information through KD. Moreover, our proposed memory-based method achieved even better PER results than the audio-lip-tongue SE model for the majority of phonemes. This observation demonstrates that the integration of memory network not only effectively recovers tongue features but also obtains higher quality tongue features through this process.

\subsection{Further Discussion on Memory-Based Audio-Lip SE Model}
\label{Section: further analysis}

\begin{figure*}[]
\begin{minipage}[b]{0.5\linewidth}
  \centering
  \centerline{\includegraphics[width=8.5cm]{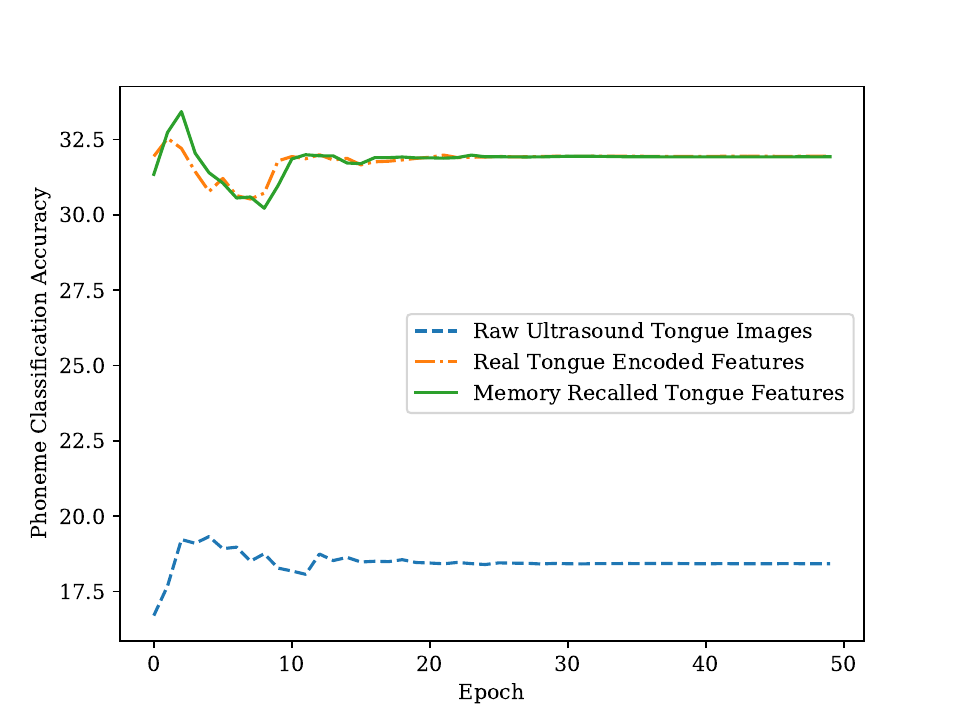}}
  \centerline{(a) Phoneme classification accuracy.}\medskip
\end{minipage}
\begin{minipage}[b]{0.5\linewidth}
  \centering
  \centerline{\includegraphics[width=8.5cm]{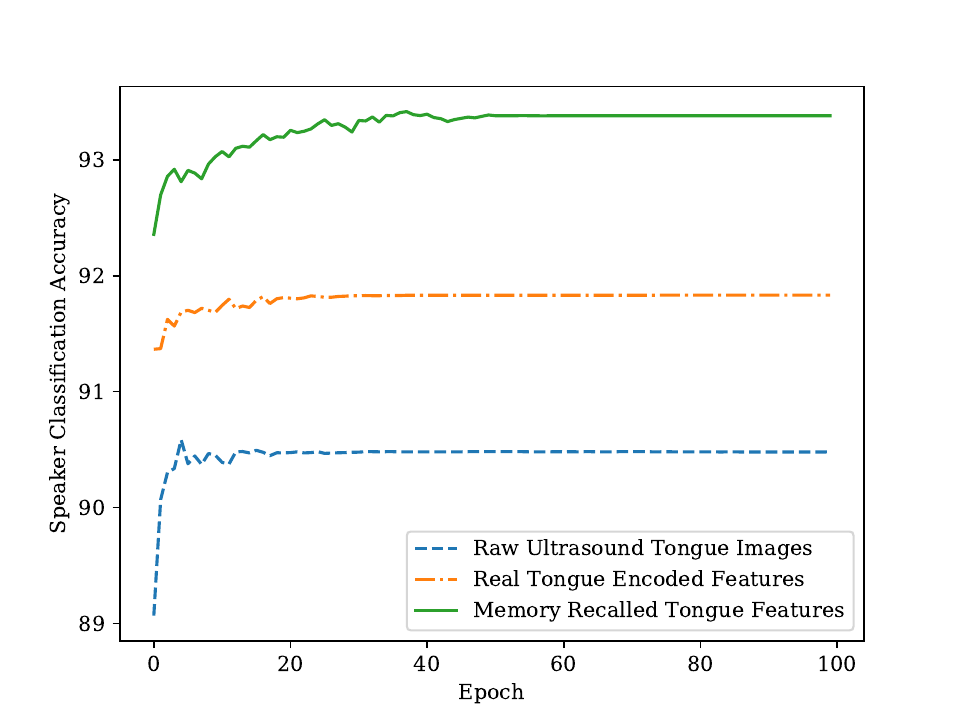}}
  \centerline{(b) Speaker classification accuracy.}\medskip
\end{minipage}
\caption{Phoneme and speaker classification accuracy using raw ultrasound tongue images, real tongue encoded features, and memory-recalled tongue features as training input. The x-axis represents the total training epochs for the classification model while the y-axis represent the classification accuracy (\%).}
\label{Figure: classification}
\end{figure*}

The results shown in Section \ref{Section: Overall Performance}, \ref{Section: generalization}, and \ref{Section: PER} have clearly shown that the proposed memory-based audio-lip SE method can achieve results that are comparable to, or even better than the audio-lip-tongue SE model. This suggests that integrating the memory network not only enables effective recovery of tongue features but also potentially yields higher quality features compared to directly encoding ultrasound tongue images. In order to delve deeper into the underlying reasons for this improvement, we conducted analytical experiments that specifically focused on the memory-based audio-lip SE model.

We constructed and trained a simple phoneme classification network and a speaker classification network, both with a 3-layer MLP architecture, using raw ultrasound tongue images, real tongue encoded features (output from the articulation convolutional block of audio-lip-tongue SE model), and memory-recalled tongue features (output recalled by the tongue-value memory of memory-based audio-lip SE model) as input, respectively. The resulting phoneme and speaker classification accuracy are illustrated in Fig.\ref{Figure: classification}. 

It can be witnessed from the results that both tongue features outperform raw ultrasound tongue images on both classification task by a large margin, indicating that the articulation convolutional block extract useful information from the input raw images. While real tongue encoded features and memory-recalled tongue features exhibited similar accuracy on the phoneme classification task, memory-recalled tongue features achieve stronger performance on the speaker classification task. This suggests that the superior performance of the memory-based audio-lip SE model may be attributed to the memory's ability to extract and utilize enhanced tongue features including cleaner and more precise speaker information. Previous studies \cite{chuang2019speaker, koizumi2020speech} have already demonstrated the substantial advantages of incorporating speaker information to enhance the performance of speech enhancement task by either integrating speaker identity embedding or employing speaker-aware features in DNN-based SE systems. These studies further confirm our hypothesis that the recalled tongue features obtained in the memory manner contain more valuable speaker information, thereby contributing to the improved performance of the model. This finding also aligns with our intuitive analysis, as the process of storing representative tongue features in a tongue value memory with a limited number of slots can be viewed as a form of quantization. As a result, useful information is retained in the memory while noisy parts of the features are discarded. Thus, the recalled tongue features obtained through memory retrieval exhibit less noise and higher quality than those directly encoded from authentic ultrasound tongue images.

\section{Conclusion}
\label{Section: Conclusion}
In conclusion, this paper introduces the incorporation of ultrasound tongue images to enhance the performance of lip-based AV-SE systems. In addition to proposing an audio-lip-tongue SE model, two methods are presented to address the challenges of acquiring ultrasound tongue images during inference. 
The first KD-based audio-lip SE method employs KD to transfer tongue knowledge from a pre-trained audio-lip-tongue SE teacher model to an audio-lip speech enhancement student model in a teacher-student learning manner, investigating the feasibility of leveraging tongue-related information without directly inputting ultrasound tongue images. The second memory-based method further integrates  a memory network into the encoder of the proposed audio-lip-tongue SE model to better model the alignment between lip and tongue modality, enabling the recovery of tongue features from readily available lip features during inference, facilitating the subsequent speech enhancement task. Experimental results demonstrate the effectiveness and robust generalization performance of both proposed audio-lip SE methods in improving speech quality and intelligibility, surpassing traditional lip-based AV-SE baselines. Additionally, the analysis of PER in ASR reveals that the incorporation of ultrasound tongue images benefits all phonemes, with tongue-related phonemes, such as palatal and velar consonants, experiencing the most significant improvement. 
Our future work will be investigating the feasibility of improving the efficiency of the existing AV-SE systems and incorporating ultrasound tongue images into other audio-visual speech processing tasks.

\bibliographystyle{IEEEtran}
\bibliography{mybib}


 





\end{document}